\documentstyle[prd,epsfig,aps]{revtex}

\begin{document}
\title{On the thickness of a mildly relativistic collisional shock wave}
\author{$^{1}$J. A. S. Lima, $^{2}$A. Kandus and $^{2}$R. Opher}
\address{$^{1}$ Departamento de F\'{\i}sica, UFRN, C.P. 1641, \\
59072-970, Natal, RN, Brazil;\\
$^{2}$Depto de Astronomia,\\
IAG-USP\\
05508-900, S\~ao Paulo, SP, Brazil}
\maketitle
\date{\today}

\begin{abstract}
We consider an imperfect relativistic fluid, in which a shock wave develops
and discuss its structure and thickness, taking into account the effects of
viscosity and heat conduction in the form of sound absorption. The junction
conditions and the nonlinear equations describing the evolution of the shock
are derived with the corresponding Newtonian limit discussed in detail. As in
the nonrelativistic regime, the thickness is inversely proportional to the 
discontinuity in the pressure. However, new terms of purely
relativistic origin are also present. In particular, for a viscous
polytropic gas, it is found that a purely viscous relativistic shock is
thicker than its nonrelativistic counterpart, while for pure heat conduction,
the contrary is true.
\end{abstract}

\section{Introduction}

The theory of relativistic shock waves was pioneered more than fifty years
ago by Taub \cite{taub}, with the related junction conditions and adiabats
further discussed by Israel \cite{I60}, Lichnerowicz \cite{LI67} and Thorne 
\cite{T73}. These results were established for a relativistic perfect simple
fluid, and since they do not involve any characteristic length
scale, the shock front was described by a mathematical surface of zero
thickness (abrupt transition). Many studies have been
made extending these works to the nonlinear regime of relativistic
hydrodynamics, as well as to ideal relativistic magnetohydrodynamics 
(e.g., see the book by A. M. Anile \cite{ANI89} and references therein).

In this work, we are interested in the shock wave theory for an imperfect
relativistic fluid. It is well known that for dissipative relativistic
fluids, for scales smaller than the dissipation scale ${\cal L}$ associated
to $\chi /c$, where $\chi $ is the viscosity or heat
conduction coefficient and $c$ is the speed of light, ordinary Navier-Stokes
formulae do not apply \cite{GER01}. We are concerned with fluid regimes
whose characteristic lengths are larger than the dissipation
scale, and therefore the classical theory for dissipative fluids may be used.

All fluids are dissipative, and a nonrelativistic shock wave
propagating in a dissipative medium cannot, in general, be considered an
abrupt transition, but instead, as a region with a finite thickness. 
Its thickness is determined 
by the dissipative coefficients (i.e., viscosity and heat conductivity) 
\cite{LL97,ZEL66}. In addition, shock waves propagating through a gas
mixture that undergoes diffusion of one component, show similar
characteristics as do those due to dissipation \cite{COW41}. This fact affects
not only the evolution of the wave but can also have important effects on processes 
that depend upon the features of the shock wave.

A pioneering study of this system was made many years ago by Koch \cite{KO65},
which showed that if the shock velocity is greater than a given critical
value, relativistic interaction of heat transfer and momentum transfer give
rise to an increase in the velocity at the upstream end of the shock layer.
The purpose of our work is to discuss several aspects of this system that were not
considered in the work by Koch, thereby providing a more complete picture of 
relativistic shock waves.

The main aim here is to derive the equation for the (generalized) Taub
curve, as well as the general expression for the thickness of a plane shock wave
in the weak relativistic regime, taking into account both the classical
dissipative mechanisms (heat conduction, bulk and shear viscosity) and
the associated sound absorption process. We shall not address the important issue
of diffusion in a relativistic fluid or
applications of our results to astrophysical processes, both of which will be 
left for future work.

The paper is structured as follows. In the next section, we review briefly
the Eckart formulation for a viscous and heat conducting relativistic
simple fluid. In section III, we derive the junction conditions, as well as the
corresponding expressions for the generalized Taub-Rankine-Hugoniot curves
for a unidimensional shock wave in a nonequilibrium regime. The general
expression for the thickness of the shock and its comparison with the
nonrelativistic limit is derived in section IV . The Newtonian limits for
the remaining expressions are discussed in the Appendix. The metric
signature is $(+,-,-,-)$.

\section{Imperfect Relativistic Fluids: The Eckart Approach}

The relativistic theory of imperfect fluids rests on two basic ideas. The
first one is the local equilibrium hypothesis (LEH). It implies that for
nonequilibrium fluids, state functions (such as entropy) depend locally on the
same set of thermodynamic variables as do equilibrium fluids. In particular, the
usual thermodynamic temperature and pressure concepts are maintained in the
relativistic nonequilibrium regime. The second idea is the existence of a
local entropy source strength (entropy variation per unit volume and unit
time), which is always nonnegative, as required by the second law of
thermodynamics. Mathematically, the LEH is represented by the Gibbs law,
whereas the entropy law takes the form of a balance equation. Using these
hypothesis in the fluid equations of motion, one finds an expression
for the entropy source strength, as well as for the constitutive
(phenomenological) relations. The perfect fluid equilibrium equations are
recovered in the limit of a vanishing entropy production rate. However,
an important point of difference in the treatment of relativistic and 
nonrelativisitic fluids by different authors should be stressed. In contrast to
the Newtonian regime, in the relativistic domain, there exists an ambiguity
related
to the possible choices of the macroscopic hydrodynamic four-velocity. In
the so-called Eckart formulation\cite{E40,W71}, the four-velocity is
directly related to the particle flux, while in Landau-Lifshitz's approach 
\cite{LL97}, it is defined by the energy flux. In principle, a general
treatment should be able to deal with any of these ``gauge'' choices\cite
{SLC02}. For simplicity and for the sake of a simpler comparison with
previous studies, in what follows, we shall adopt the Eckart formulation.

The thermodynamic state of a relativistic simple fluid is characterized by
an energy-momentum tensor $T^{{\alpha }{\beta }}$, a particle current $%
N^{\alpha }$, and an entropy current $S^{\alpha }$. The fundamental equations
are expressed by the conservation laws (particles and energy-momentum) and
the entropy flux equation,
\begin{equation}
N_{,\mu }^{\mu }=0,\,\,\,T_{,\nu }^{\mu \nu }=0,\,\,\,S_{,\mu }^{\mu }\geq 0,
\label{eck-a}
\end{equation}
where $N^{\mu }$ is the particle flux, $T^{\mu \nu }$ the stress tensor,
and $S^{\mu }$ is the entropy flux (comma denotes space-time derivatives).
In the Eckart frame, the particle flux and stress tensor can be written as 
\cite{W71,SLC02} 
\begin{equation}
N^{\mu }=nu^{\mu },  \label{eck-b}
\end{equation}
\begin{equation}
T^{\mu \nu }=\mu u^{\mu }u^{\nu }-ph^{\mu \nu }+\pi h^{\mu \nu
}+c^{-1}(q^{\mu }u^{\nu }+q^{\nu }u^{\mu })+\Pi ^{\mu \nu },  \label{eck-c}
\end{equation}
with the entropy flux given by 
\begin{equation}
S^{\mu }=nk_{B}\sigma u^{\mu }-\frac{q^{\mu }}{T}.  \label{eck-d}
\end{equation}
The quantities
$n$, $\rho $, $p$, $\sigma $, $T$, and $k_{B}$ are
the particle concentration, energy density, pressure, specific entropy (per
particle), temperature, and the Boltzmann constant, respectively. The 
hydrodynamic 4-velocity $%
u^{\mu}$ is normalized according to $u^{\mu }u_{\mu }=1$. The tensor 
\begin{equation}
h^{\mu \nu }=g^{\mu \nu }-u^{\mu }u^{\nu }
\end{equation}
is the usual projector onto the local rest space of $u^{\alpha }$. The
irreversible fluxes, $\pi $, $q^{\mu }$, and $\Pi ^{\mu \nu }$, are defined by

\begin{eqnarray}  \label{CE}
&&\pi =\zeta \theta \,,  \label{CE1} \\
&&q^{\mu }=\kappa h^{\mu \nu }(T,_{\nu }-Ta_{\nu })\,,  \label{CE2} \\
&&\Pi ^{\mu \nu }=\eta \left( h^{\mu \alpha }u^{\nu },_{\alpha }+h^{\nu
\alpha }u^{\mu },_{\alpha }-\frac{2}{3}\theta h^{\mu \nu }\right)\,,
\label{CE3}
\end{eqnarray}
where $\kappa $, $\zeta $ and $\eta $ are the classical phenomenological
coefficients (thermal conductivity, bulk and shear viscosity), and $a_{\nu
}=u_{\nu ,\alpha }u^{\alpha }$ is the four acceleration. The bulk viscosity
stress, $\pi $, represents an irreversible negative pressure, and $\theta $
is the scalar of expansion (divergence of 4-velocity). The heat flux $q^{\mu
}$ is orthogonal to the 4-velocity, i.e., $q^{\mu }u_{\mu }=0$, whereas the
shear-viscosity tensor $\Pi ^{\mu \nu }$, is symmetric, trace free, and
space-like.

For completeness, we recall that all dissipative fluxes $\pi $, $q^{\mu }$,
and $\Pi ^{\mu \nu }$, as well as their space-time derivatives, are of first
order of smallness in the equilibrium deviations. This is also true of the
space-time derivatives of the reversible thermodynamic quantities $n$, $\rho$
, $p$, $\sigma $, and $T$. However, the source of entropy, i.e., the
divergence of the entropy flux 
\begin{equation}
{S^{\mu }}_{;\mu }={\frac{\pi ^{2}}{\zeta T}} - {\frac{q^{\alpha }q_{\alpha }
}{\kappa T}}+{\frac{\Pi ^{\alpha \beta }\Pi _{\alpha \beta }}{2\eta T}}\,,
\end{equation}
is a quantity of second order of smallness. In what follows, we write the
4-velocity as $u^{\mu }=\gamma \left( 1,\vec{v}/c\right) $, where $\gamma
=\left( 1-v^{2}/c^{2}\right) ^{-1/2}$ is the Lorentz factor.

\section{Taub curves and entropy density change}

We now consider the junction conditions for a plane shock wave
in a relativistic imperfect fluid and use them to derive the
generalized Taub curve, as well as the associated entropy density change 
for weak shocks. The thickness
of the shock wave due to the presence of viscosity and thermal conduction
and taking into account the acoustic damping is, then, derived.

\subsection{Junction Conditions}

In an ideal fluid, the relativistic junction conditions are defined by the 
continuity equation for the particle current $N^{x}$ and the momentum
and energy fluxes, i.e., the $xx$ and $0x$ components of the energy-momentum
tensor: 
\begin{equation}
\left[ N^{x}\right] =0,\quad c\left[ T^{0x}\right] =0,\quad \left[ T^{xx}
\right] =0\,.  \label{junc-a}
\end{equation}
Square brackets denote the difference between the values of any of the
mentioned quantities at large distances in front of the shock and 
inside it. 
We denote side 1 as the (far) upstream side. Choosing the spatial component of the
four velocity along the $x$-axis, it follows that $u^{0}=\gamma $, $
u^{x}=\gamma v^{x}/c$. The non-null components of the projector tensor
are: $h^{00}=-\left(u^{x}\right) ^{2}$, $h^{0x}=-\gamma u^{x}$ and $
h^{xx}=-\gamma ^{2}$. For convenience, the density particle current will be
expressed as $j=n\gamma v$. In this way, the conservation of the $0x$ and $xx
$ components of the energy-momentum tensor take the form: 
\begin{eqnarray}
&&n\gamma \frac{v}{c}\equiv \frac{j}{c}=\frac{j_{1}}{c}\,,  \label{junc-d} \\
&&\left( \text{{\sf w}}\gamma -\text{{\sf w}}_{1}\gamma _{1}\right) =-\left(
\zeta +\frac{4}{3}\eta \right) \frac{j}{c}\frac{\gamma }{n^{3}}\left[\frac{j 
}{c^{2}}\frac{\partial _{t}n}{\gamma n}+\partial _{x}n\right]\,,
\label{junc-i} \\
&&+\frac{\kappa }{c}\left( \gamma ^{2}+\frac{j^{2}}{c^{2}}\frac{1}{n^{2}}
\right) \left\{ \frac{1}{cn}\partial _{t}T+\frac{c}{j}\gamma \partial _{x}T- 
\frac{T}{cn^{2}}\left[ \frac{j}{n}\frac{\partial _{x}n}{\gamma }+\partial
_{t}n\right] \right\}\,,  \nonumber
\end{eqnarray}
\begin{eqnarray}
&&\frac{j^{2}}{c^{2}}\left( \frac{\text{{\sf w}}}{n}-\frac{\text{{\sf w}}
_{1} }{n_{1}}\right) +\left( p-p_{1}\right) \left. =\right. -\left( \zeta +
\frac{4 }{3}\eta \right) \frac{j}{c}\frac{\gamma ^{2}}{n^{2}}\left[ \frac{j}{
c^{2}} \frac{\partial _{t}n}{\gamma n}+\partial _{x}n\right]  \label{junc-j}
\\
&&+2\frac{\kappa }{c}\gamma \frac{j^{2}}{c^{2}}\frac{1}{n}\left\{ \frac{1}{
cn }\partial _{t}T+\frac{c}{j}\gamma \partial _{x}T-\frac{T}{cn^{2}}\left[ 
\frac{j}{n}\frac{\partial _{x}n}{\gamma }+\partial _{t}n\right] \right\}\,, 
\nonumber
\end{eqnarray}
where we have introduced the specific enthalpy (per particle), 
\begin{equation}
{\sf w}=\frac{\mu +p}{n}\,,  \label{junc-e}
\end{equation}
and used $u_{,0}^{0}=u^{x}\partial _{t}u^{x}/c\gamma ,\quad
u_{,x}^{0}=u^{x}\partial _{x}u^{x}/\gamma $. It has been assumed that
at large distances from the shock, the flux is uniform, i.e., all
gradients vanish.

\subsection{Generalized Taub curve}

In order to obtain the expression for the change in the entropy across the
shock, we follow a procedure similar to that adopted by Thorne \cite
{T73}. First we multiply (\ref{junc-j}) by $\left( \text{{\sf w}}/n+\text{
{\sf w}}_{1}/n_{1}\right) $ and then combine the result with
$j^{2}=n_{1}^{2}\left(
u_{1}^{x}\right) ^{2}c^{2}=n^{2}\left( u^{x}\right) ^{2}c^{2}$, obtaining 
\begin{eqnarray}
&&\text{{\sf w}}^{2}u^{x2}-\text{{\sf w}}_{1}^{2}u_{1}^{x2}+\left(
p-p_{1}\right) \left( \frac{\text{{\sf w}}}{n}+\frac{\text{{\sf w}}_{1}}{
n_{1}}\right) \left. =\right. -\left( \zeta +\frac{4}{3}\eta \right) \gamma
^{2}\left( \frac{\text{{\sf w}}}{n}+\frac{\text{{\sf w}}_{1}}{n_{1}}\right) 
\frac{j}{cn^{2}}\left[ \frac{j}{nc^{2}}\frac{\partial _{t}n}{\gamma }
+\partial _{x}n\right]  \nonumber \\
&&+2\frac{\kappa }{c}\gamma \frac{j^{2}}{c^{2}}\frac{1}{n}\left( \frac{\text{
{\sf w}}}{n}+\frac{\text{{\sf w}}_{1}}{n_{1}}\right) \left\{ \frac{1}{n} 
\frac{1}{c}\partial _{t}T+\frac{c}{j}\gamma \partial _{x}T-\frac{T}{cn^{2}}
\left[ \frac{j}{n}\frac{\partial _{x}n}{\gamma }+\partial _{t}n\right]
\right\}\,.  \label{junc-k}
\end{eqnarray}

Multiplying (\ref{junc-i}) by $\left( \text{{\sf w}}\gamma +\text{{\sf w
}}_{1}\gamma_{1}\right) $, we get 
\begin{eqnarray}
&&\left( \text{{\sf w}}^{2}\gamma ^{2}-\text{{\sf w}}_{1}^{2}\gamma
_{1}^{2}\right) =-\left( \zeta +\frac{4}{3}\eta \right) \left( \text{{\sf w}}
\gamma +\text{{\sf w}}_{1}\gamma _{1}\right) \frac{j}{c}\frac{\gamma }{n^{3}}
\left[ \frac{j}{c^{2}}\frac{\partial _{t}n}{\gamma n}+\partial _{x}n\right]
\label{junc-l} \\
&&+\frac{\kappa }{c}\left( \gamma ^{2}+\frac{j^{2}}{c^{2}}\frac{1}{n^{2}}
\right) \left( \text{{\sf w}}\gamma +\text{{\sf w}}_{1}\gamma _{1}\right)
\left\{ \frac{1}{cn}\partial _{t}T+\frac{c}{j}\gamma \partial _{x}T-\frac{T}{
cn^{2}}\left[ \frac{j}{n}\frac{\partial _{x}n}{\gamma }+\partial
_{t}n\right] \right\} \,.  \nonumber
\end{eqnarray}
Finally, subtracting (\ref{junc-l}) from (\ref{junc-k}) and using $
\gamma ^{2}=1+\left( u^{x}\right) ^{2}$, we obtain 
\begin{eqnarray}
&&\text{{\sf w}}^{2}-\text{{\sf w}}_{1}^{2}\left. =\right. \left(
p-p_{1}\right) \left( \frac{\text{{\sf w}}}{n}+\frac{\text{{\sf w}}_{1}}{
n_{1}}\right)  \label{junc-m} \\
&&+\left( \zeta +\frac{4}{3}\eta \right) \gamma ^{2}\gamma _{1}\text{{\sf w}}
_{1}\frac{1}{n^{2}}\frac{j}{c}\left[ \frac{1}{\gamma _{1}n_{1}}-\frac{1}{
\gamma n}\right] \left[ \frac{j}{c^{2}}\frac{\partial _{t}n}{\gamma n}
+\partial _{x}n\right]  \nonumber \\
&&+\kappa \left[ \text{{\sf w}}\gamma +\text{{\sf w}}_{1}\gamma _{1}-2\frac{
j^{2}}{c^{2}}\text{{\sf w}}_{1}\gamma _{1}\frac{\gamma }{n}\left( \frac{1}{
\gamma _{1}n_{1}}-\frac{1}{\gamma n}\right) \right]  \nonumber \\
&&\times \left\{ \frac{\gamma }{j}\partial _{x}T+\frac{1}{nc^{2}}\partial
_{t}T-\frac{T}{c^{2}n^{2}}\left[ \frac{j}{n}\frac{\partial _{x}n}{\gamma }
+\partial _{t}n\right] \right\}\,.  \nonumber
\end{eqnarray}
Equation (\ref{junc-m}), together with the definition of $j/c$ [Eq. (
\ref{junc-d})], are the generalized Taub junction conditions for a plane shock
wave in an imperfect relativistic simple fluid.

\subsection{Weak shock wave: entropy density change}
We now consider the weak shock case, i.e., that for which all
discontinuities are small. This means that differences,such as $V-V_{1}$, $
p-p_{1}$, etc., between the values in front of the transition layer
and inside it are small. Thus differentiation with respect to $x$ or $ct$ 
increases the
order of smallness by one, i.e., $dV/dx$ is a quantity of second order of 
smallness. From (\ref{junc-m}), we see that the term involving the viscosity
coefficients are of third order, while for heat conduction, the terms
proportional to $\left( \text{{\sf w}}\gamma +\text{{\sf w}}_{1}\gamma
_{1}\right) $ are of second order. The number particle conservation law can
be written as 
\begin{equation}
\frac{u^{\alpha}{n_{,\alpha}}}{n}=-\theta \,,  \label{junc-n}
\end{equation}
and considering that to lowest order, the temperature gradient satisfies 
\cite{SLC02} 
\begin{equation}
\frac{u^{\alpha }{T_{,\alpha }}}{T}=-\left( \frac{\partial p}{\partial \mu }
\right) _{n}\theta \,,  \label{junc-o}
\end{equation}
the enthalpy density change, given by (\ref{junc-m}), can be expressed as 
\begin{eqnarray}
\text{{\sf w}}^{2}-\text{{\sf w}}_{1}^{2} &=&\left( p-p_{1}\right) \left( 
\text{{\sf w}}/n+\text{{\sf w}}_{1}/n\right)  \label{weak-a} \\
&&+\kappa \left( \text{{\sf w}}\gamma +\text{{\sf w}}_{1}\gamma _{1}\right)
\left\{ \frac{\gamma }{j}\partial _{x}T-\frac{j}{\gamma c^{2}n^{2}}\partial
_{x}T+\frac{T}{\gamma nc}\left[ 1-\left( \frac{\partial p}{\partial \mu }
\right) _{n}\right] \theta \right\} \,.  \nonumber
\end{eqnarray}
Note that the term proportional to $T$ is proportional to $c^{-2}$ through
the dependence of $\theta $ on the four velocity and time derivative (cf.
eq. (\ref{junc-n})).

In the dissipationless regime, the resulting expression for the Taub adiabat
is formally very similar to the Newtonian expression, as can be seen in
Refs. \cite{LL97} and \cite{T73}. For an imperfect relativistic fluid, the change 
in entropy in the transition layer is also of second order in the
pressure, just as it is in the nonrelativistic case. However, as one may see from
(\ref{weak-a}) three new purely relativistic terms come into play. In the
Appendix, we show that eq. (\ref{weak-a}) yields the Newtonian expression
previously found in the literature (e.g., Ref. \cite{LL97}).

To find the expression for the difference in the entropy density values
far upstream and in the transition layer, we follow
a standard procedure \cite{T73} and develop ${\sf w}/n$ around its upstream
value in powers of $\left( p-p_{1}\right) $. We write the first law of
thermodynamics as $d${\sf w}$=dp/n+Td{\sf s}$, where ${\sf s}$ is the
entropy per particle and then multiply by {\sf w}, using the
development of ${\sf w}/n$. Keeping the zeroth order in {\sf w}$T$ in the
second term and integrating, we get 
\begin{eqnarray}
\text{{\sf w}}^{2}-\text{{\sf w}}_{1}^{2} &=&2\text{{\sf w}}_{1}T_{1}\left(
s-s_{1}\right) +2\frac{\text{{\sf w}}_{1}}{n_{1}}\left( p-p_{1}\right)
\label{weak-e} \\
&&+\left[ \frac{\partial }{\partial p}\left( \frac{\text{{\sf w}}}{n}\right)
\right] _{s,1}\left( p-p_{1}\right) ^{2}+\frac{1}{3}\left[ \frac{\partial
^{2}}{\partial p^{2}}\left( \frac{\text{{\sf w}}}{n}\right) \right]
_{s,1}\left( p-p_{1}\right) ^{3}\,.  \nonumber
\end{eqnarray}
As the derivatives of $T$ and $n$ are already of second order, we consider $
\left( \text{{\sf w}}\gamma +\text{{\sf w}}_{1}\gamma _{1}\right) \simeq 2$ 
{\sf w}$_{1}\gamma _{1}$ in eq. (\ref{weak-a}). With this approximation and
comparing with (\ref{weak-e}), we obtain the entropy density change: 
\begin{equation}
{\sf s}-{\sf s}_{1}\simeq \frac{\kappa }{T_{1}}\gamma _{1}\left\{ \frac{
\gamma }{j}\partial _{x}T-\frac{j}{\gamma c^{2}n}\partial _{x}T+\frac{T}{
\gamma nc}\left[ 1-\left( \frac{\partial p}{\partial \mu }\right)
_{n}\right] \theta \right\} \,.  \label{weak-f}
\end{equation}
Therefore, as in the nonrelativistic case, the entropy density change is
proportional to the heat conduction coefficient. The nonrelativistic 
limit of this expression is trivial and coincides with the
known expression \cite{LL97}.

\section{Shock wave thickness}

Relativistic or nonrelativistic shocks are described by an evolving non
linear wave. On the other hand, waves propagating in a viscous, heat
conducting medium are damped. This fact can be phenomenologically
described by an extra imaginary term in the dispersion relationship for the
wave, i.e., by writing $\omega \simeq v_{s}k-icLk^{2}$, where $\omega $ is
the frequency, $k$ the wavenumber, $v_{s}$ the sound speed, and $L$ is the
absorption length (see Refs. \cite{LL97,WAPJ71}). The equation we are
seeking must be of the form \cite{LL97}: 
\begin{equation}
\left( \frac{\partial }{\partial t}-v_{s}\frac{\partial }{\partial x}\right)
f-v_{s}\alpha _{p}f\frac{\partial }{\partial x}f=cL\frac{\partial ^{2}}{
\partial x^{2}}f\,,  \label{shock-a}
\end{equation}
where $f$ is a suitable function that describes the wave profile.
To obtain this equation, we shall follow a two-step procedure: 
we first find the nonlinear term (in the next subsection) and then proceed to
find the quasi-acoustic damping contribution (in the subsequent subsection).

\subsection{Nonlinear term in shock waves}

In order to find the nonlinear contribution, we need only to consider 
equations for an ideal fluid. We 
consider now the local reference frame, in which the medium is at rest
(comoving frame), and let $\delta v$ be a unidimensional velocity
perturbation. We have (cf. Ref. \cite{W71})

\begin{equation}
\frac{\partial }{\partial t}\mu +\delta v\frac{\partial }{\partial x}\mu
+\left( \mu +p\right) \left[ \frac{1}{c^{2}}\delta v\frac{\partial }{
\partial t}\delta v+\frac{\partial }{\partial x}\delta v\right] =0 \,,
\label{nonl-a}
\end{equation}

\begin{equation}
\frac{\left( \mu +p\right) }{c^{2}}\left[ \frac{\partial \delta v}{\partial t
}+\delta v\frac{\partial }{\partial x}\delta v\right] +\left[ \frac{\delta v
}{c^{2}}\frac{\partial p}{\partial t}+\frac{\partial }{\partial x}p\right]
=0\,,  \label{nonl-b}
\end{equation}
Expanding $\mu$ and $p$, we have 
\begin{equation}
\mu =\mu _{0}+\frac{c^{2}}{v_{s}^{2}}\delta p+\frac{1}{2}\left( \frac{
\partial ^{2}\mu }{\partial p^{2}}\right) _{s}\delta p^{2}\,,  \label{nonl-c}
\end{equation}
\begin{equation}
p=p_{0}+\delta p\,,  \label{nonl-d}
\end{equation}
where $\mu _{0}$ and $p_{0}$ are the background values. For a
wave propagating to the left, we can write $\delta v=-\left(
c^{2}/v_{s}\right) \delta p/\left( \mu +p\right) $ and using 
$\partial /\partial t=v_{s}\partial /\partial x$ in the
second order terms, (\ref{nonl-b}) can be written as 
\begin{equation}
\frac{1}{c^{2}}\left( \mu _{0}+p_{0}\right) \frac{\partial \delta v}{
\partial t}+\frac{\partial }{\partial x}\delta p=\frac{2}{\left( \mu
_{0}+p_{0}\right) }\delta p\frac{\partial }{\partial x}\delta p\,
\label{nonl-g}
\end{equation}
and Eq. (\ref{nonl-a}) as
\begin{equation}
\frac{c^{2}}{v_{s}^{2}}\frac{\partial }{\partial t}\delta p+\left( \mu
_{0}+p_{0}\right) \frac{\partial }{\partial x}\delta v=v_{s}\left\{ \frac{2}{%
\left( \mu _{0}+p_{0}\right) }\frac{c^{4}}{v_{s}^{4}}-\left( \frac{\partial
^{2}\mu }{\partial p^{2}}\right) _{s}\right\} \delta p\frac{\partial }{%
\partial x}\delta p \,.  \label{nonl-i}
\end{equation}

Deriving (\ref{nonl-g}) with respect to $x$ and (\ref{nonl-i}) with
respect to $t$ and subtracting the resulting expressions, we get 
\begin{equation}
\left( \frac{1}{v_{s}}\frac{\partial }{\partial t}-\frac{\partial }{\partial
x}\right) \left( \frac{1}{v_{s}}\frac{\partial }{\partial t}+\frac{\partial 
}{\partial x}\right) \delta p=\left( \frac{v_{s}}{c^{2}}\alpha _{1}\frac{
\partial }{\partial t}-\alpha _{2}\frac{\partial }{\partial x}\right) \left[
\delta p\frac{\partial }{\partial x}\delta p\right]\,,
\end{equation}
where 
\begin{equation}
\alpha _{1}=\left\{ \frac{2}{\left( \mu _{0}+p_{0}\right) }\frac{c^{4}}{
v_{s}^{4}}-\left( \frac{\partial ^{2}\mu }{\partial p^{2}}\right)
_{s}\right\}\,,  \label{nonl-k}
\end{equation}
\begin{equation}
\alpha _{2}=\frac{2}{\left( \mu _{0}+p_{0}\right) }\,.  \label{nonl-l}
\end{equation}
Substituting $\partial /\partial t=v_{s}\partial /\partial x$ and
eliminating $\partial /\partial x$ in both terms, we get 
\begin{equation}
\left( \frac{1}{v_{s}}\frac{\partial }{\partial t}-\frac{\partial }{\partial
x}\right) \delta p-\alpha _{p}\delta p\frac{\partial }{\partial x}\delta
p=0\,,  \label{nonl-n}
\end{equation}
where we define
\begin{equation}
\alpha _{p}\equiv \frac{1}{2}\left( \frac{v_{s}^{2}}{c^{2}}\alpha
_{1}-\alpha _{2}\right) =\frac{1}{2}\left[ \frac{2}{\left( \mu +p\right) } 
\frac{c^{2}}{v_{s}^{2}}-\frac{v_{s}^{2}}{c^{2}}\left( \frac{\partial ^{2}\mu 
}{\partial p^{2}}\right) _{s}-\frac{2}{\left( \mu +p\right) }\right]\,.
\label{nonl-o}
\end{equation}
This equation has the same form as does the nonrelativistc one 
(see Ref. \cite{LL97}), with
$\left( \mu +p\right) $ replacing the rest mass density, the
relativistic energy density derived twice with respect to the pressure,
and the ratio of the sound speed to the light speed appearing explicitly. The
last term in the square brackets is a purely relativistic correction.

\subsection{Dissipative term of the shock wave equation}

The acoustic relativistic damping length required by the complete nonlinear
equation of a shock wave was derived in another context by Weinberg
\cite{WAPJ71}. We refer the interested reader to this work in order to see
details of the derivation. Here, we just quote the final expression, 
\begin{equation}
cL=\frac{1}{2\left( \mu +p\right) }\left\{ c^{2}\left( \zeta +\frac{4}{3}
\eta \right) +\kappa \left( \mu +p\right) \left( \frac{1}{C_{v}}-\frac{1}{
C_{p}}\right) +\kappa T\left[ \frac{v_{s}^{2}}{c^{2}}-\frac{2}{C_{v}}\left( 
\frac{\partial p}{\partial T}\right) _{n}\right] \right\}\,,  \label{nonl-p}
\end{equation}
observing that Weinberg's expression is recovered when 
$C_{v}\left( C_{p}\right) =nc_{v}\left( nc_{p}\right) $ and $c=1$. The above 
expression is the same as the nonrelativistic
result (see Ref. \cite{LL97}), with $\left( \mu +p\right) $ replacing
the density of the rest mass and a relativistic correction
proportional to $\kappa T$ appears explicitly.

\subsection{Solving for the thickness}

The complete
equation of the evolution of a shock wave is obtained by adding to eq. (\ref
{nonl-n}) a term proportional to the second derivative with respect to 
$x $, which takes into account the dissipation. The final equation is then  
\begin{equation}
\left( \frac{\partial }{\partial t}-v_{s}\frac{\partial }{\partial x}\right)
\delta p-v_{s}\alpha _{p}\delta p\frac{\partial }{\partial x}\delta p=cL 
\frac{\partial ^{2}\delta p}{\partial x^{2}}\,.  \label{thick-a}
\end{equation}
Following the usual analysis \cite{LL97}, we assume that $\delta p$ has the
following dependence:
\begin{equation}
\delta p=\delta p\left( \xi \right) ,\qquad \xi =x+v_{w}t  \label{thick-b}
\end{equation}
where $v_{w}$ is the velocity of the wave. With this solution, Eq.(\ref{thick-a}) 
becomes
\begin{equation}
\frac{d}{d\xi }\left[ \left( v_{w}-v_{s}\right) \delta p-\frac{1}{2}
v_{s}\alpha _{p}\delta p^{2}-cL\frac{d}{d\xi }\delta p\right] =0 \,.
\label{thick-c}
\end{equation}
The solution to Eq.(\ref{thick-a}) is then \cite{LL97} 
\begin{equation}
p=\frac{1}{2}\left( p_{1}+p_{2}\right) +\frac{1}{2}\left( p_{2}-p_{1}\right)
\tanh \frac{\left( p_{2}-p_{1}\right) \left( x+v_{w}t\right) }{4\left(
c/v_{s}\right) \left( L/\alpha _{p}\right)}\,,  \label{thick-h}
\end{equation}
where $p_{1}$ is the pressure far upstream and $p_{2}$,the pressure far
downstream.
In the reference frame where the shock is at rest, we have for the
pressure variation 
\begin{equation}
p-\frac{1}{2}\left( p_{1}+p_{2}\right) =\frac{1}{2}\left( p_{2}-p_{1}\right)
\tanh \left( \frac{x}{\bigtriangleup }\right)\,,  \label{thick-i}
\end{equation}
where we have defined the ``thickness'' of the shock by 
\begin{equation}
\bigtriangleup =\frac{4cL}{v_{s}\alpha _{p}\left( p_{2}-p_{1}\right)}\,.
\label{thick-j}
\end{equation}
We see that this expression is identical in form to the nonrelativistic
one and proportional to the inverse of the pressure difference. The
relativistic corrections are contained in the factors $\alpha _{p}$ and $cL$.

\subsection{Analysis of the thickness}

In this subsection, we shall estimate the effect of the relativistic
corrections to see if they increase or decrease the shock thickness. We shall
examine them in the weak relativistic limit. We then write (see Appendix) 
\begin{equation}
cL=\Lambda _{NR}-\lambda =\Lambda _{NR}\left[ 1-\frac{\lambda }{\Lambda _{NR}
}\right]  \label{comp-a}
\end{equation}
and 
\begin{equation}
\alpha _{p}=v_{s}^{2}\tilde{\alpha}_{NR}-\eta =v_{s}^{2}\tilde{\alpha}
_{NR}\left[ 1-\frac{\eta }{v_{s}^{2}\tilde{\alpha}_{NR}}\right]\,,
\label{comp-b}
\end{equation}
with $\Lambda _{NR}$, $\lambda ,\tilde{\alpha}_{NR}$ and $\eta $ given in the
Appendix. The nonrelativistic expression for the shock thickness is 
\cite{LL97} $\delta =4a/\tilde{\alpha}_{NR}\left( p_{2}-p_{1}\right) $ where $
a=\Lambda _{NR}/v_{s}^{3}$. We must evaluate 
\begin{equation}
\frac{\bigtriangleup }{\delta }=\frac{cL\tilde{\alpha}_{NR}}{v_{s}\alpha
_{p}a}\simeq 1+\frac{\eta }{v_{s}^{2}\tilde{\alpha}_{NR}}-\frac{\lambda }{
\Lambda _{NR}}\,,  \label{comp-c}
\end{equation}
where the semi-equality holds for the weak relativistic case. Using
the expressions in the Appendix, we find 
\begin{eqnarray}
\frac{\bigtriangleup }{\delta } &=&1+\frac{2v_{s}^{2}}{\rho c^{2}}\frac{
\left\{ 1+\frac{\left( \varepsilon +p\right) }{\rho v_{s}^{2}}+\frac{
v_{s}^{2}}{2}\rho \left( \frac{\partial ^{2}\varepsilon }{\partial p^{2}}
\right) _{s}\right\} }{\left[ \frac{2}{\rho }-v_{s}^{4}\left( \frac{\partial
^{2}\rho }{\partial p^{2}}\right) _{s}\right] }  \label{comp-d} \\
&&-\frac{1}{\rho c^{2}}\frac{\left[ \left( \zeta +\frac{4}{3}\eta \right)
\left( \varepsilon +p\right) +\frac{\kappa T}{c_{v}}\left( \frac{\partial p}{
\partial T}\right) _{n}\right] }{\left[ \left( \zeta +\frac{4}{3}\eta
\right) +\frac{\kappa }{2}\left( \frac{1}{c_{v}}-\frac{1}{c_{p}}\right)
\right]}\,.  \nonumber
\end{eqnarray}
For a more direct comparison of the relativistic thickness
with the standard Newtonian result, we consider a polytropic gas and
evaluate the above expression in two special cases: with viscosity alone and
with thermal conduction alone.

\subsubsection{Polytropic gas}

In a classical polytropic gas, the energy density and
enthalpy density are given by $\varepsilon
=c_{v}T=p/\left( \Gamma -1\right) $ and $w=c_{p}T=\Gamma p/\left( \Gamma
-1\right) $, where $\Gamma =c_{p}/c_{v}=const$, respectively. Hence $\left(
\partial
^{2}\varepsilon /\partial p^{2}\right) _{s}=0$ and $\left( \partial
p/\partial T\right) _{n}=c_{v}\left( \Gamma -1\right) $ and $
1/c_{v}-1/c_{p}=\left( \Gamma -1\right) ^{2}T/\Gamma p$. Replacing these
formulae in the classical expressions for the internal energy $\varepsilon$
in eq. (\ref{comp-d}), we get 
\begin{equation}
\frac{\bigtriangleup }{\delta }=1+\frac{v_{s}^{2}}{c^{2}}\frac{2\gamma }{
\left( \Gamma ^{2}-1\right) }-\frac{1}{\rho c^{2}}\frac{\left[ \left( \zeta
+ \frac{4}{3}\eta \right) \frac{\Gamma p}{\left( \Gamma -1\right) }+\kappa
T\left( \Gamma -1\right) \right] }{\left[ \left( \zeta +\frac{4}{3}\eta
\right) +\frac{\kappa }{2}\frac{\left( \Gamma -1\right) ^{2}T}{\Gamma p}
\right]}\,.  \label{comp-e}
\end{equation}

\paragraph{Only viscosity.}

If thermal conduction is absent, we find 
\begin{equation}
\frac{\bigtriangleup }{\delta }=1+\frac{v_{s}^{2}}{c^{2}}\frac{1}{\left(
\Gamma +1\right)} \,.  \label{comp-f}
\end{equation}
We see that, in this case, the relativistic shock is thicker than the
nonrelativistic one, with the increment proportional to the sound
speed.

\paragraph{Only heat conduction.}

If viscosity is absent, we obtain 
\begin{equation}
\frac{\bigtriangleup }{\delta }=1-2\frac{v_{s}^{2}}{c^{2}}\frac{1}{\left(
\Gamma ^{2}-1\right)}\,.  \label{comp-g}
\end{equation}
In this case, the relativistic shock is thinner than its nonrelativistic
counterpart and the correction is again proportional to the sound speed.

\subsection{Entropy density change}

With the expression for the pressure given by (\ref{thick-i}), we can
express the entropy change as a function of the pressure discontinuity. We
begin by writing explicitly the derivatives in the expression for $\theta $,
namely eq. (\ref{junc-n}) and replacing the time derivative with 
$\partial /\partial t=v_{s}\partial
/\partial x$. Using $dT/dx=\left(
\partial T/\partial p\right) _{s}dp/dx+\left( \partial T/\partial s\right)
_{p}ds/dx\simeq \left( \partial T/\partial p\right) _{s}dp/dx$ and $
dn/dx=\left( \partial n/\partial p\right) _{s}dp/dx+\left( \partial
n/\partial s\right) _{p}ds/dx\simeq \left( \partial n/\partial p\right)
_{s}dp/dx$ in Eq.(\ref{weak-f}) and evaluating $dp/dx$
from (\ref{thick-i}), we obtain the following expression for the entropy
density change in a reference system in which the shock is at rest:

\begin{eqnarray}
{\sf s}-{\sf s}_{1} &\simeq &\frac{\kappa }{T_{1}}\gamma _{1}\left\{ \left[ 
\frac{1}{j}-\frac{j}{\gamma c^{2}n}\right] \left( \frac{\partial T}{\partial
p}\right) _{s}-\frac{T\left( v_{s}+v\right) }{\gamma n^{2}c^{2}}\left[
1-\left( \frac{\partial p}{\partial \mu }\right) _{n}\right] \left( \frac{
\partial n}{\partial p}\right) _{s}\right\}  \nonumber \\
&\times &\frac{v_{s}\alpha _{p}}{8cL}\frac{\left( p_{2}-p_{1}\right) ^{2}}{
\cosh ^{2}\left( x/\bigtriangleup \right) }\,,  \label{thick-m}
\end{eqnarray}
where the factor $c$ in $cL$ does not add an extra power in the speed of 
light (see Appendix).
In the nonrelativistic case \cite{LL97}, the entropy reaches a
maximum inside the shock \cite{ZEL66} and is of second order in the pressure
discontinuity.

\section{Conclusion}

In this paper, we have extended previous studies of shock waves done in the
nonrelativistic domain to the weak relativistic case. Considering
dissipative relativistic fluids in the range of validity of the
Navier-Stokes-Fourier theory \cite{GER01}, we have obtained expressions for
the entropy density change and the shock thickness that coincide in form with
nonrelativistic ones. In each of the factors in the equations, purely
relativistic corrections appear explicitly. We studied the
expression for the shock thickness for a polytropic gas and analyzed the
effect of corrections in two important limits defined by the
presence of viscosity or heat conduction. When only heat conduction is taken
into account, the relativistic shock is thinner than for the nonrelativistic
case. This result can be understood by observing that heat conducting
fluids can develop ``thermal discontinuities'' \cite{LL97,ZEL66}, i.e., they
allow for discontinuities in the velocity, pressure and density of the fluid
flow, while the temperature remains constant. On the other hand, when only
viscosity is present, the shock thickness is larger than for its nonrelativistic
counterpart and, hence, the tendency to erase singularities is stronger in
the relativistic limit than in the Newtonian one. This difference in the
effect of the relativistic corrections can also be understand, as follows.
Viscosity provides the mechanism to convert a portion of kinetic energy of
the gas flowing into the discontinuity into heat. This conversion is
equivalent to the transformation of the energy of ordered motion of gas
molecules into energy of random motion by the dissipation of molecular
motion. In this respect, heat conduction has an indirect effect on the
conversion process since it only participates in the transfer of the energy
of random motion of the molecules from one point to another, but does not
directly affect the ordered motion. The corresponding relativistic
corrections seem to amplify these effects.

A comment on the entropy change is in order. When the pre-shock gas has a low
temperature, we are in the strictly Newtonian limit and, in this sense, the fact
that the entropy density change reduces to its Newtonian analog is equivalent
to requiring that the theory has the correct low-speed limit. This fact
contains no new information. But when the pre-shock fluid has relativistic
internal speeds, the shock weakness does not imply a Newtonian propagation
velocity of the shock and, hence, this case is not covered by the Newtonian
treatment. In this sense, the result that we have obtained, that the entropy
change
still reduces to the Newtonian expression is new and potentially interesting.

Finally, it should be mentioned that although first-order theories are successful 
in revealing the physics underlying a large class of phenomena, they 
present some experimental and theoretical drawbacks. In its classical
version, the linear constitutive equations (\ref{CE1})-(\ref{CE3}) are not
adequate at high frequencies or short wave lengths, as manifested in
experiments on ultrasound propagation in rarefied gases and on neutron
scattering in liquids \cite{Jou89}. In additon, they also allow for the
propagation
of perturbations with arbitrarily high speeds, which although unsatisfactory
on classical grounds, is completely unacceptable from a relativistic point
of view. Furthermore, they do not have a well-posed Cauchy problem and their
equilibrium states are not stable. Several authors have formulated
relativistic second-order theories which circumvent these defficiencies\cite
{Dixon78,Israel76,Pavon82,Hiscock83}. In a forthcoming paper, we intend to
extend our considerations to this class of theories.

{\bf Acknowledgments:} The authors thank G. Medina-Tanco for
helpful discussions and the anonymous referee for invaluable
comments and criticisms on this paper. The authors would also like to
thank the Brazilian agencies FAPESP (Proc.Nos. 00/06695-0 and 00/06770-2),
FINEP (Pronex No. 41.96.0908.00) and CNPq for partial support.
A.K. thanks FAPESP (Proc.No.01/07748-3) for a post-doctoral fellowship.

\section{Appendix}

In this appendix, we obtain the nonrelativistic limits of the 
magnitudes discussed in this paper. We begin with the nonrelativistic
limit of $\gamma ^{2}\left( \mu +p\right) $, neglecting the
pressure since  $nmc^{2}\gg p$ in the nonrelativistic limit. Thus,
$\gamma ^{2}\left( \mu +p\right)
\rightarrow \mu =\gamma ^{2}nmc^{2}+\gamma ^{2}\varepsilon $, where $
\varepsilon $ is the internal energy density, i.e., the energy associated with
internal degrees of freedom. In the limit of small velocities, $
nm\rightarrow \frac{\rho }{\gamma }$, where $\rho $ is the mass density and,
therefore,  $\gamma ^{2}nmc^{2}\rightarrow \gamma \rho c^{2}=\left( 1-v^{2}/c^{2}
\right) ^{-1/2}\rho c^{2}\simeq \rho c^{2}+(1/2)\rho v^{2}$. Taking $
\gamma =1$ in the expression for the internal energy density, we obtain the
desired limit: $\gamma ^{2}\left( \mu +p\right) \rightarrow \rho
c^{2}+(1/2)\rho v^{2}+\rho \epsilon $, where $\epsilon $ is the internal
energy per particle.

\subsection{Nonrelativistic limit of the Taub curve}

At the limit $c\rightarrow \infty $ in Eq.(\ref{weak-a}), the
terms in the square brackets can be
neglected. Thus, 
\begin{equation}
{\sf w}^{2}-{\sf w}_{1}^{2}=\left( p-p_{1}\right) \left( \frac{{\sf w}}{n}+ 
\frac{{\sf w}_{1}}{n_{1}}\right) +\frac{\kappa }{j}\left( {\sf w}+{\sf w}
_{1}\right) \partial _{x}T \,,  \label{ap2-e}
\end{equation}
or
\begin{equation}
{\sf w}^{2}-\left( p-p_{1}\right) \frac{{\sf w}}{n}-\frac{\kappa }{j}{\sf w}
\partial _{x}T={\sf w}_{1}^{2}+\left( p-p_{1}\right) \frac{{\sf w}_{1}}{
n_{1} }+\frac{\kappa }{j}{\sf w}_{1}\partial _{x}T \,.  \label{ap2-f}
\end{equation}
Taking the square-root,  
\begin{equation}
{\sf w}\left[ 1-\left( p-p_{1}\right) \frac{1}{{\sf w}n}-\frac{\kappa }{j 
{\sf w}}\partial _{x}T\right] ^{1/2}={\sf w}_{1}\left[ 1+\left(
p-p_{1}\right) \frac{1}{{\sf w}_{1}n_{1}}+\frac{\kappa }{j{\sf w}_{1}}
\partial _{x}T\right] ^{1/2} \,.  \label{ap2-g}
\end{equation}
Assuming that $\left( p-p_{1}\right)$ and $\partial _{x}T$ are small,
we have
\begin{equation}
{\sf w}\left( 1-\frac{1}{2}\left( p-p_{1}\right) \frac{1}{{\sf w}n}-\frac{1}{
2}\frac{\kappa }{j{\sf w}}\partial _{x}T\right) ={\sf w}_{1}\left( 1+\frac{1 
}{2}\left( p-p_{1}\right) \frac{1}{{\sf w}_{1}n_{1}}+\frac{1}{2}\frac{\kappa 
}{j{\sf w}_{1}}\partial _{x}T\right) \,  \label{ap2-h}
\end{equation}
or, rearranging terms, 
\begin{equation}
{\sf w}-{\sf w}_{1}=\frac{1}{2}\left( p-p_{1}\right) \left( \frac{1}{n}+ 
\frac{1}{n_{1}}\right) +\frac{\kappa }{j}\partial _{x}T \,,  \label{ap2-i}
\end{equation}
which is the standard nonrelativistic expression for the Taub Curve \cite
{LL97}.

\subsection{Nonrelativistic limit of the shock thickness}

The nonrelativistic limit of the shock thickness is derived from Eqs (34) 
and (35), the expressions for nonrelativistic $\alpha _{p}$ and $L$,
respectively. Using the expression 
for $\gamma^2\left(\mu +p\right) $, derived in the
introduction to the Appendix, we rewrite eq. (\ref{nonl-o}) for $\alpha
_{p} $ as 
\begin{equation}
\alpha _{p}=\frac{v_{s}^{2}}{2}\left[ \frac{2}{\rho \left( 1+\varepsilon
/\rho c^{2}+p/\rho c^{2}\right) }\frac{1}{v_{s}^{4}}-\left[ \frac{\partial
^{2}}{\partial p^{2}}\left( \rho +\frac{\varepsilon }{c^{2}}\right) \right]
_{s}-\frac{2}{v_{s}^{2}\left( \rho c^{2}+\varepsilon +p\right) }\right] \,.
\label{ap2-j}
\end{equation}
In the weak relativistic limit, we obtain 
\begin{equation}
\alpha _{p}=\alpha_{NR}-\eta \,,  \label{ap2-k}
\end{equation}
with 
\begin{equation}
\alpha _{NR}=\frac{v_{s}^{2}}{2}\left[ \frac{2}{\rho v_{s}^{4}}-\left( \frac{
\partial ^{2}\rho }{\partial p^{2}}\right) _{s}\right] =v_{s}^{2}\tilde{
\alpha}_{NR} \,,  \label{ap2-l}
\end{equation}
\begin{equation}
\eta =\frac{1}{\rho c^{2}}\left\{ 1+\frac{\left( \varepsilon +p\right) }{
\rho v_{s}^{2}}+\frac{v_{s}^{2}}{2}\rho \left( \frac{\partial
^{2}\varepsilon }{\partial p^{2}}\right) _{s}\right\}\,.  \label{ap2-m}
\end{equation}
It is
convenient to express the dispersion relationship as $k=\gamma \omega/v_s
+i\gamma ^{2}cL\omega ^{2}/v_s^3$, where $L$ is defined in Eq.(35). Using 
$\left(\mu +p\right) /c^{2}=\rho
+\left(\varepsilon +p\right)/c^{2}$, in the weak relativistic limit, we have 
\begin{equation}
cL=\Lambda _{NR}-\lambda =\Lambda _{NR}\left[ 1-\frac{\lambda }{\Lambda
_{NR} }\right] \,,  \label{ap2-p}
\end{equation}
where 
\begin{equation}
\Lambda _{NR}=\frac{1}{2\rho }\left[ \left( \zeta +\frac{4}{3}\eta \right) + 
\frac{\kappa }{2}\left( \frac{1}{c_{v}}-\frac{1}{c_{p}}\right) \right]\,,
\label{ap2-q}
\end{equation}
\begin{equation}
\lambda =\frac{1}{2\rho ^{2}c^{2}}\left[ \left( \zeta +\frac{4}{3}\eta
\right) \left( \varepsilon +p\right) +\frac{\kappa T}{2}\frac{2}{c_{v}}
\left( \frac{\partial p}{\partial T}\right) _{n}\right]\,,  \label{ap2-r}
\end{equation}
with $c_{v}=C_{v}/\rho $, the specific heat per unit mass.

Using Eqs.(\ref{ap2-k})-(\ref{ap2-r}) in eq. (\ref
{thick-j}) and taking the limit $c\rightarrow \infty $, we obtain the standard
$\alpha _{p}$ expression for the thickness of a nonrelativistic 
shock \cite{LL97}: 
\begin{equation}
\delta =\frac{8aV^{2}}{\left( \partial ^{2}V/\partial p^{2}\right)
_{s}\left( p_{2}-p_{1}\right)} \,.  \label{ap2-s}
\end{equation}

\end{document}